\documentclass{aa}  

\usepackage{graphicx}
\usepackage{txfonts}
\usepackage{hyperref}
\usepackage{ulem}
\hypersetup{
    colorlinks=true,
    linkcolor=blue,
    filecolor=magenta,  
    citecolor=blue,
    urlcolor=cyan,
    pdfpagemode=FullScreen,
    }

\def\setsymbol#1#2{\expandafter\def\csname #1\endcsname{#2}}
\def\getsymbol#1{\csname #1\endcsname}

\newbox\tablebox    \newdimen\tablewidth
\def\leaderfil{\leaders\hbox to 5pt{\hss.\hss}\hfil}
\def\tablenote#1 #2\par{\begingroup \parindent=0.8em
    \abovedisplayshortskip=0pt\belowdisplayshortskip=0pt
    \noindent
    $$\hss\vbox{\hsize\tablewidth \hangindent=\parindent \hangafter=1 \noindent
    \hbox to \parindent{$^#1$\hss}\strut#2\strut\par}\hss$$
    \endgroup}

\def\L2{\ifmmode L_2\else $L_2$\fi}

\def\DeltaT{\ifmmode \Delta T\else $\Delta T$\fi}
\def\deltat{\ifmmode \Delta t\else $\Delta t$\fi}
\def\fknee{\ifmmode f_{\rm knee}\else $f_{\rm knee}$\fi}
\def\Fmax{\ifmmode F_{\rm max}\else $F_{\rm max}$\fi}
\def\solar{\ifmmode{\rm M}_{\mathord\odot}\else${\rm M}_{\mathord\odot}$\fi}
\def\Msolar{\ifmmode{\rm M}_{\mathord\odot}\else${\rm M}_{\mathord\odot}$\fi}
\def\Lsolar{\ifmmode{\rm L}_{\mathord\odot}\else${\rm L}_{\mathord\odot}$\fi}
\def\inv{\ifmmode^{-1}\else$^{-1}$\fi}
\def\mo{\ifmmode^{-1}\else$^{-1}$\fi}
\def\sup#1{\ifmmode ^{\rm #1}\else $^{\rm #1}$\fi}
\def\expo#1{\ifmmode \times 10^{#1}\else $\times 10^{#1}$\fi}
\def\,{\thinspace}
\def\lsim{\mathrel{\raise .4ex\hbox{\rlap{$<$}\lower 1.2ex\hbox{$\sim$}}}}
\def\gsim{\mathrel{\raise .4ex\hbox{\rlap{$>$}\lower 1.2ex\hbox{$\sim$}}}}

\def\simprop{\mathrel{\raise .4ex\hbox{\rlap{$\propto$}\lower 1.2ex\hbox{$\sim$}}}}
\def\deg{\ifmmode^\circ\else$^\circ$\fi}
\def\pdeg{\ifmmode $\setbox0=\hbox{$^{\circ}$}\rlap{\hskip.11\wd0 .}$^{\circ}
          \else \setbox0=\hbox{$^{\circ}$}\rlap{\hskip.11\wd0 .}$^{\circ}$\fi}
\def\arcs{\ifmmode {^{\scriptstyle\prime\prime}}
          \else $^{\scriptstyle\prime\prime}$\fi}
\def\arcm{\ifmmode {^{\scriptstyle\prime}}
          \else $^{\scriptstyle\prime}$\fi}
\newdimen\sa  \newdimen\sb
\def\parcs{\sa=.07em \sb=.03em
     \ifmmode \hbox{\rlap{.}}^{\scriptstyle\prime\kern -\sb\prime}\hbox{\kern -\sa}
     \else \rlap{.}$^{\scriptstyle\prime\kern -\sb\prime}$\kern -\sa\fi}
\def\parcm{\sa=.08em \sb=.03em
     \ifmmode \hbox{\rlap{.}\kern\sa}^{\scriptstyle\prime}\hbox{\kern-\sb}
     \else \rlap{.}\kern\sa$^{\scriptstyle\prime}$\kern-\sb\fi}
\def\ra[#1 #2 #3.#4]{#1\sup{h}#2\sup{m}#3\sup{s}\llap.#4}
\def\dec[#1 #2 #3.#4]{#1\deg#2\arcm#3\arcs\llap.#4}
\def\deco[#1 #2 #3]{#1\deg#2\arcm#3\arcs}
\def\rra[#1 #2]{#1\sup{h}#2\sup{m}}

\def\dots{\relax\ifmmode \ldots\else $\ldots$\fi}
\def\WHzsr{\ifmmode $W\,Hz\mo\,sr\mo$\else W\,Hz\mo\,sr\mo\fi}
\def\mHz{\ifmmode $\,mHz$\else \,mHz\fi}
\def\GHz{\ifmmode $\,GHz$\else \,GHz\fi}
\def\mKs{\ifmmode $\,mK\,s$^{1/2}\else \,mK\,s$^{1/2}$\fi}
\def\muKs{\ifmmode \,\mu$K\,s$^{1/2}\else \,$\mu$K\,s$^{1/2}$\fi}
\def\muKRJs{\ifmmode \,\mu$K$_{\rm RJ}$\,s$^{1/2}\else \,$\mu$K$_{\rm RJ}$\,s$^{1/2}$\fi}
\def\muKHz{\ifmmode \,\mu$K\,Hz$^{-1/2}\else \,$\mu$K\,Hz$^{-1/2}$\fi}
\def\MJysr{\ifmmode \,$MJy\,sr\mo$\else \,MJy\,sr\mo\fi}
\def\MJysrmK{\ifmmode \,$MJy\,sr\mo$\,mK$_{\rm CMB}\mo\else \,MJy\,sr\mo\,mK$_{\rm CMB}\mo$\fi}
\def\microns{\ifmmode \,\mu$m$\else \,$\mu$m\fi}

\def\muK{\ifmmode \,\mu$K$\else \,$\mu$\hbox{K}\fi}
\def\microK{\ifmmode \,\mu$K$\else \,$\mu$\hbox{K}\fi}
\def\muW{\ifmmode \,\mu$W$\else \,$\mu$\hbox{W}\fi}
\def\kms{\ifmmode $\,km\,s$^{-1}\else \,km\,s$^{-1}$\fi}
\def\kmsMpc{\ifmmode $\,\kms\,Mpc\mo$\else \,\kms\,Mpc\mo\fi}
\providecommand{\sorthelp}[1]{}

\begin{document}

   \title{Planck data revisited: low-noise synchrotron polarisation maps from the WMAP and Planck space missions}
\titlerunning{Low-noise synchrotron polarisation maps from WMAP and Planck}
   \author{Jacques Delabrouille
          \thanks{\email{delabrouille@apc.in2p3.fr}} \inst{\ref{cpb},\ref{lbl}}}

\institute{CNRS-UCB International Research Laboratory, Centre Pierre Bin\'etruy, IRL 2007, CPB-IN2P3, Berkeley, CA 94720, USA \label{cpb} \and Lawrence Berkeley National Laboratory, 1 Cyclotron Road, Berkeley, CA 94720, USA \label{lbl}}

\date{Received ; accepted }

% context, aims, methods, results, conclusions
\abstract{Observations of cosmic microwave background polarisation, essential for probing a potential phase of inflation in the early universe, suffer from contamination by polarised emission from the Galactic interstellar medium.}
{This work combines existing observations from the WMAP and Planck space missions to make a low-noise map of polarised synchrotron emission that can be used to clean forthcoming CMB observations.}
{We combine WMAP K, Ka and Q maps with Planck LFI 30~GHz and 44~GHz maps using weights that near-optimally combine the observations as a function of sky direction, angular scale, and polarisation orientation.}
{We publish well-characterised maps of synchrotron $Q$ and $U$ Stokes parameters at $\nu = 30$~GHz and 1 degree angular resolution. A statistical description of uncertainties is provided with Monte-Carlo simulations of additive and multiplicative errors.}
{Our maps are the most sensitive full-sky maps of synchrotron polarisation to date, and are made available to the scientific community on a dedicated web site.}


\maketitle

\section{Introduction}

The cosmic microwave background (CMB), relic radiation emitted when electrons and nuclei first combined to form neutral atoms, has been transformational for understanding the origin, evolution, and matter and energy content of the Universe we live in. Observations from a series of space-borne \citep{1992ApJ...396L...1S,1994ApJ...420..439M,2013ApJS..208...20B,2020A&A...641A...1P},
balloon-borne \citep[e.g.,][]{2000ApJ...545L...5H,2002ApJ...571..604N,2003A&A...399L..19B}, 
and ground-based \citep[e.g.,][]{2014JCAP...04..014D,2016PhRvL.116c1302B,2017JCAP...06..031L,2017ApJ...848..121P,2018ApJ...852...97H} CMB
experiments, complemented by other cosmological probes \citep[e.g.,][]{1999ApJ...517..565P,2005ApJ...633..560E,2017MNRAS.470.2617A,2022PhRvD.105b3520A}, have led to a standard cosmological scenario, inflationary $\Lambda$CDM, in which an expanding universe is filled today with about 30\% pressureless matter and 70\% dark energy.
In this paradigm, galaxies and clusters of galaxies form from initial density perturbations generated during an early phase of rapid expansion, dubbed cosmic inflation.
The fast expansion in the inflationary phase expands tiny local quantum fluctuations of the spacetime metric to supra-horizon scales, which later-on re-enter the horizon and seed the gravitational collapse of structures observable today \citep{1983PhRvD..28..679B}. While appealing, the inflationary hypothesis still requires observational confirmation, as well as the identification of the correct model among a plethora of viable alternatives \citep{2014PDU.....5...75M}. 

In many inflationary models, initial perturbations of the spacetime metric comprise, in addition to a scalar term corresponding to initial density perturbations responsible for the formation of the large scale structures, a tensor term corresponding to primordial gravitational waves. These tensor modes of metric perturbation are expected to imprint, on the CMB, tiny fluctuations of polarisation of the B-mode type (of odd parity), while density perturbations generate, at the time of CMB emission, only E-mode type polarisation (of even parity). For this reason, the detection of primordial $B$
modes of CMB polarisation is considered an essential next step in observational cosmology \citep{2016ARA&A..54..227K}. If detected with an angular power spectrum matching theoretical prediction, they would be a ``smoking gun'' for cosmic inflation. This motivates the deployment of ambitious next-generation CMB observatories, on the ground and in space, dedicated to the detection of primordial CMB polarisation B-modes \citep{2018SPIE10708E..07H,2019JCAP...02..056A,2019arXiv190704473A,2018JCAP...04..014D,2019arXiv190210541H,2023PTEP.2023d2F01L}.

The amplitude of primordial gravitational waves, parameterised by the tensor-to-scalar ratio $r$, is already constrained by existing CMB observations to $r \leq 0.036$ and $r \leq 0.032$ at 95\% CL, in two recent independent analyses combining Planck and BICEP/Keck data \citep{2022arXiv220316556B,2022PhRvD.105h3524T}. At this level and below, over most of the sky, for most of the harmonic scales, and at all observing frequencies, CMB B-mode fluctuations are fainter than fluctuations of microwave polarised emission originating from the interstellar medium (ISM) of our own Milky Way. Two processes of ISM emission are known to contribute a significant amount of microwave polarization. Elongated galactic dust grains, aligned preferentially perpendicularly to the local Galactic magnetic field, polarise the light from background stars parallel to the local Galactic magnetic field \citep{1951ApJ...114..206D}, and emit polarised microwave emission perpendicular to the magnetic field. Polarised dust emission in the microwave and sub-millimeter has been detected at 353 GHz by the Archeops stratospheric balloon \citep{2004A&A...424..571B}, and in several frequency bands by the high frequency instrument aboard the Planck space mission \citep{2020A&A...641A..11P,2020A&A...641A...4P}. At lower frequency, energetic electrons in the ionised ISM spiralling in the Galactic magnetic field generate synchrotron emission that is highly polarised. This polarised synchrotron emission has been mapped over the full sky, although with limited signal-to-noise ratio at high galactic latitude, by the WMAP space mission \citep{2007ApJ...665..355K}, and by the low frequency instrument (LFI) aboard the Planck satellite \citep{2016A&A...594A..25P,2020A&A...641A...4P}.

\begin{table*}[t]
    \centering
    \begin{tabular}{|c||c|c|c|c|c|}
        \hline
        Channel / Frequency & $\beta_s = -2.9$ & $\beta_s = -3.0$ & $\beta_s = -3.1$ & $\beta_s = -3.2$ & $\beta_s = -3.3$ \\
        \hline
        \hline
        WMAP K & 2.22 & 2.28 & 2.34 & 2.42 & 2.49 \\
        LFI 30~GHz & 1.20 & 1.21 & 1.22 & 1.23 & 1.24 \\
        30~GHz & 1.00 & 1.00 & 1.00 & 1.00 & 1.00 \\
        WMAP Ka & 0.77 & 0.76 & 0.75 & 0.75 & 0.74 \\
        WMAP Q & 0.42 & 0.41 & 0.40 & 0.39 & 0.37 \\
        LFI 44~GHz & 0.34 & 0.33 & 0.32 & 0.31 & 0.29 \\
        \hline
    \end{tabular}
    \caption{Synchrotron frequency scaling (in K$_{\rm CMB}$ units) for various values of the spectral index $\beta_s$. 
    Relative errors on the frequency scalings are seen to be of the order of a few per cent (10\% at most). This will induce errors of at most a few per cent of the signal in the final maps. 
    }
    \label{tab:scalings}
\end{table*}

Future sub-orbital CMB observations, such as planned with the BICEP array \citep{2018SPIE10708E..07H}, the Simons observatory \citep{2019JCAP...02..056A}, AliCPT \citep{2020SPIE11453E..2AS}, or with the future CMB-S4 experiment \citep{2019arXiv190704473A}, must monitor the contamination of CMB observations by Galactic dust and Galactic synchrotron polarisation. To that effect, one may envisage using existing data from WMAP and Planck to help model and subtract polarised emission from the ISM in ground-based CMB observations. Template maps for polarised dust and synchrotron emission have been produced with a component separation pipeline from the final Planck space mission data \citep{2020A&A...641A..11P}, and made available to the scientific community in the Planck space mission legacy archive at ESA.\footnote{\url{http://www.cosmos.esa.int/web/planck/pla}} These maps, however, suffer from limitations, some of which can be addressed making better use of existing data.

The main limitation of the published WMAP and Planck polarised synchrotron maps is the level of instrumental noise. In particular, maps are strongly noise-dominated in regions of sky of lowest Galactic foreground emission, which would be the main targets for upcoming ground-based observations. This comes from the limited sensitivity of existing surveys, and from the fact that published maps do not best combine all existing observations: the latest Planck published synchrotron maps \citep{2020A&A...641A..11P}, for instance, use only Planck data. 

In this paper, we combine WMAP and Planck LFI data in a simple yet near-optimal way, to produce well-characterised maps of the $Q$ and $U$ Stokes parameters of  ``Planck-Revisited'' polarised synchrotron emission at 30~GHz, that can be used for foreground cleaning in upcoming CMB observations with ground-based experiments. 
%We compare our resulting synchrotron polarisation map with other similar public data products.
While our paper was in preparation, \cite{2023arXiv231013740W} have combined Planck and WMAP observations to produce 30~GHz maps of polarised synchrotron emission amplitude with noise levels lower than those of the WMAP K channel alone, using a conceptually different approach. Data products are part of the Cosmoglobe first data release (Cosmoglobe DR1). In that  analysis, \cite{2023arXiv231013740W} use a pipeline that first corrects for a model of systematic residuals in WMAP data and for calibration uncertainties in Planck HFI data, and then fit for a multi-component model of sky emission in multi-frequency maps. However, our focussed pipeline has specific advantages, which we discuss in this paper. In particular, our data products have better signal-to-noise ratio in regions of low foreground contamination, and their effective angular resolution is better characterised. The increased sensitivity is of specific interest for ground-based experiments targeting primordial B-mode detection in patches of sky with low foreground emission, such as the Simons Observatory \citep{2019JCAP...02..056A},  various geneerations of BICEP/Keck \citep{2018SPIE10708E..07H,2020JLTP..199..976S} and AliCPT \citep{2020SPIE11453E..2AS}.

The rest of the paper is organised as follows. In Sec.~\ref{sec:input}, we describe the input WMAP and Planck observations that are used for this work. In Sec.~\ref{sec:method}, we describe the data pipeline that is used to combine those observations to make a single map of polarised synchrotron emission at a frequency of 30~GHz. Section~\ref{sec:results} gives and discusses the main results and data products made available to the scientific community. We conclude in Sec.~\ref{sec:conclusion}.

\section{Input observations and modeling assumptions}
\label{sec:input}

WMAP, launched in June 2001 by NASA, has observed the full microwave sky in five frequency bands, centred from 23\,GHz to 94\,GHz. The Planck space mission has observed the sky in 9 frequency bands centred at frequencies ranging from 27\,GHz to 857\,GHz.

In this paper, we use DR5 (9-year) WMAP data release polarisation maps for the K, Ka, and Q bands \citep{2013ApJS..208...20B}, and Planck DR4 LFI 30\,GHz and 44\,GHz data \citep{2020A&A...643A..42P}. Contamination by polarised dust emission at these frequencies is strongly sub-dominant as compared to either synchrotron emission or instrumental noise. 
We do not use the WMAP V and W bands, which have low synchrotron signal-to-noise ratio over the largest fraction of sky, nor the 70\,GHz Planck LFI map or any of the Planck HFI maps, which provide little additional information for mapping polarised synchrotron specifically, and may contain non-negligible polarised dust emission. 
However, we make use of the CMB temperature and polarisation maps obtained with the SMICA method \citep{2003MNRAS.346.1089D,2008ISTSP...2..735C} to correct the low-frequency maps from a sub-dominant CMB polarisation contamination.

To first order, synchrotron emission $s(\nu)$ scales in frequency proportionally to a power of frequency, i.e. for each sky pixel
\begin{equation}
    s(\nu) \propto \nu^{\beta_s}.
\end{equation}
The spectral index $\beta_s$ depends on the spectral energy distribution of the relativistic electrons, and is hence theoretically supposed to depend on the physical properties of the region of emission. 

In the frequency range of interest, a simple power law with fixed $\beta_s \simeq -3.1$ (when the emission is expressed in antenna temperature (K$_{\rm RJ}$) units) is a good approximation, but there is evidence of slight variations $\delta \beta_s \simeq \pm 0.1$ as a function of the direction in the sky  \citep{2018A&A...618A.166K,2023MNRAS.519.3504D,2023arXiv231013740W}. 
More generally, the power-law scaling approximation can break down at higher frequency, e.g. because of a depletion of high-energy electrons, leading to a possible steepening of the frequency dependence. At low frequency, below $\nu \sim 5$~GHz, Faraday rotation of polarisation and line-of-sight integration leads to depolarisation that this power law invalid for the $Q$ and $U$ Stokes parameters. 
Several authors have attempted to map the synchrotron spectral index using existing data sets, but results are noisy and uncertain \citep[see, e.g.,][for recent results]{2018A&A...618A.166K,2021A&A...646A..69F,2022MNRAS.517.2855D,2022ApJ...936...24W,2023MNRAS.519.3504D,2023arXiv231013740W}.
Spectral index maps obtained using intensity observations in the frequency domain of interest suffer from contamination by dust ``anomalous microwave emission'' (AME) and free-free emission, which are hard to separate from synchrotron by reason of lack of observing frequencies. Spectral index maps obtained from polarisation data, negligibly impacted by AME and free-free which are at most very faintly polarised, are either contaminated by large scale systematic effects due to intensity to polarisation leakage, or suffer from Faraday rotation effects and/or poor signal to noise ratio.

For the work presented here, we assume a fixed synchrotron spectral index $\beta_s = -3.1$. Table \ref{tab:scalings} gives the frequency scalings of synchrotron for several values of the spectral index, normalised to unity at 30~GHz. Relative differences from 23~GHz to 44~GHz are in the few per cent range. Errors made by using a fixed spectral index are thus of the order of a few per cent of the total emission. Further discussion of the impact of this choice is deferred to Sec.~\ref{sec:specind}.

\section{Method}
\label{sec:method}

\begin{figure}
    \centering
    \includegraphics[width=0.49\textwidth]{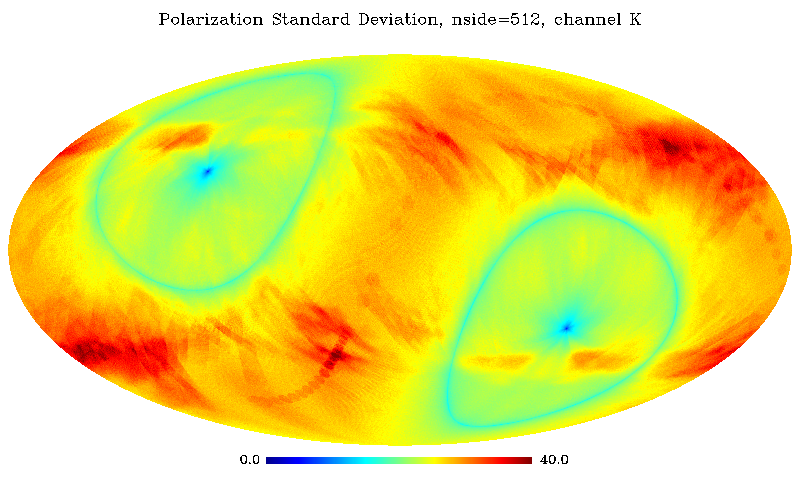}
    \includegraphics[width=0.49\textwidth]{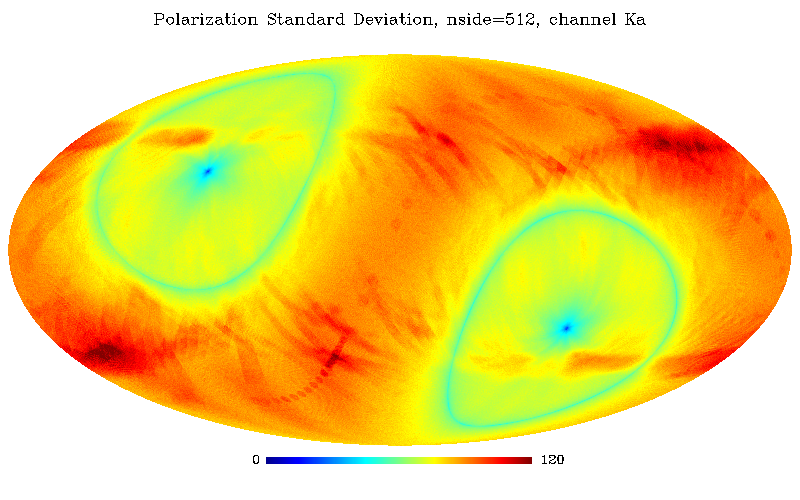}
    \includegraphics[width=0.49\textwidth]{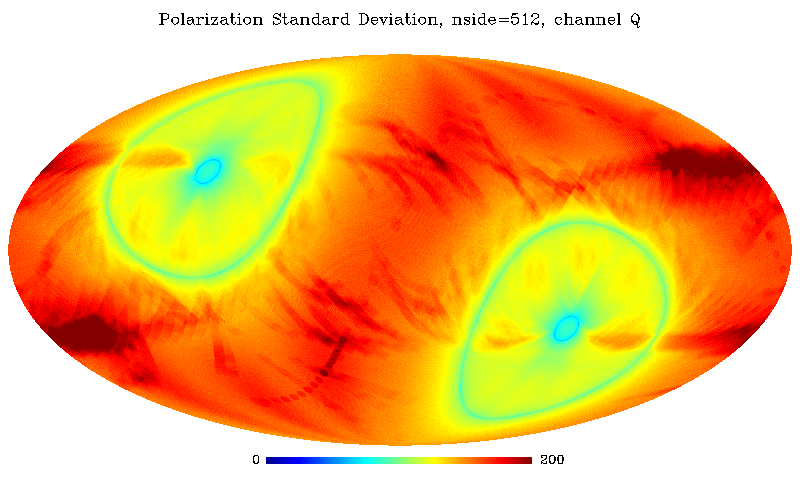}
    \caption{Noise standard deviation maps for the three WMAP channels used in this work (K, Ka and Q), re-scaled to equivalent 30~GHz noise assuming a synchrotron spectral index of $-3.1$. Units are in $\mu$K$_{\rm cmb}$. Each map displays the average of the standard deviation of the Q and U Stokes parameter maps.}
    \label{fig:noise-levels-WMAP}
\end{figure}
\begin{figure}
    \centering
    \includegraphics[width=0.49\textwidth]{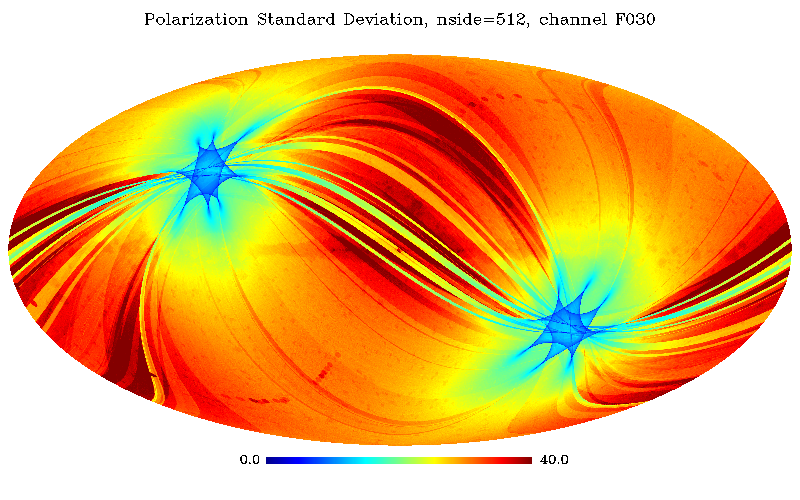}
    \includegraphics[width=0.49\textwidth]{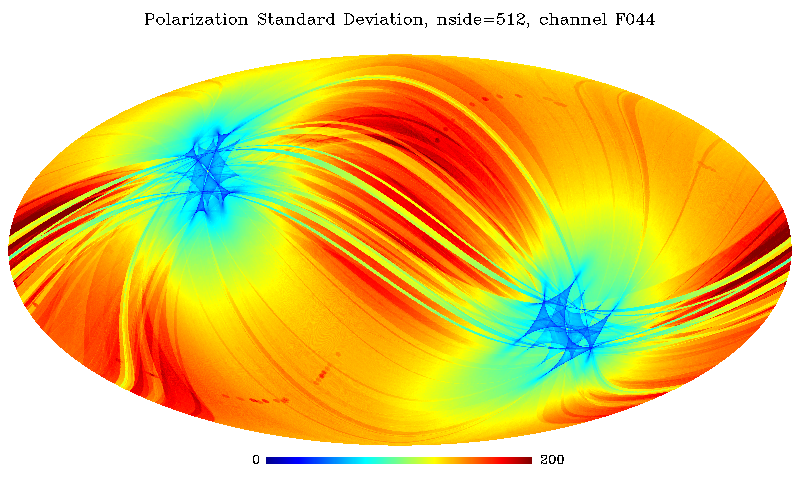}
    \caption{Noise standard deviation maps for the two Planck LFI channels used in this work (30~GHz and 44~GHz), re-scaled to equivalent 30~GHz noise assuming a synchrotron spectral index of $-3.1$. Units are in $\mu$K$_{\rm cmb}$. Each map displays the average of the standard deviation of the Q and U Stokes parameter maps. Note, here and in Fig.~\ref{fig:noise-levels-WMAP}, the difference in color scale for the different frequency channels, and the variablilty of the noise level across the sky. The map depth varies by factors of a few across the sky, and the lowest noise regions are different between the maps. This calls for pixel-dependent weights for map combination.}
    \label{fig:noise-levels-LFI}
\end{figure}

The present works aims at combining WMAP and Planck synchrotron observations for increased signal to noise ratio. For both WMAP and Planck, the noise level strongly depends on the sky pixel considered (Fig. \ref{fig:noise-levels-WMAP} and \ref{fig:noise-levels-LFI}).

\begin{figure}
    \centering
    \includegraphics[width=0.49\textwidth]{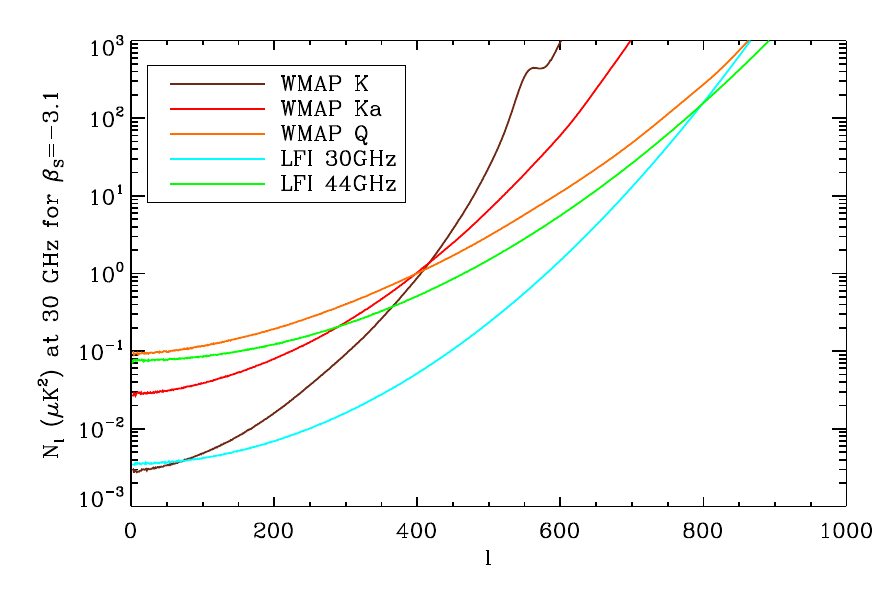}
    \includegraphics[width=0.49\textwidth]{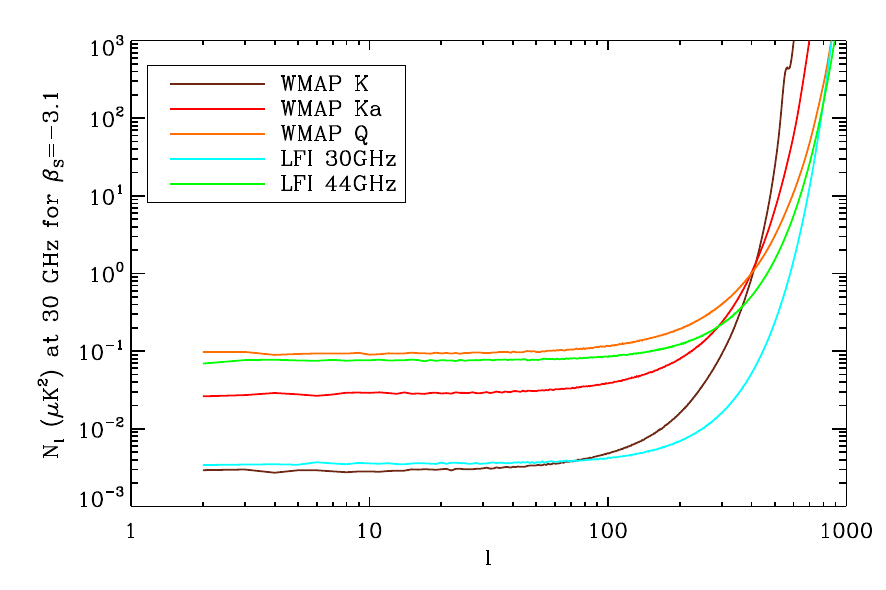}
    \caption{Beam-corrected noise spectra for the WMAP and Planck channels used in this work, re-scaled to equivalent 30~GHz noise assuming a synchrotron spectral index of $-3.1$. The relative sensitivity of the different channels strongly depends on the harmonic scale $\ell$. On the largest scales, the WMAP K band and the LFI 30~GHz channels have comparable mean synchrotron sensitivity (with, however, a strong pixel dependence, as seen in Figs.~\ref{fig:noise-levels-WMAP} and~\ref{fig:noise-levels-LFI}. At $\ell \simeq 800$, however the WMAP Q and LFI 44~GHz are about as sensitive as the LFI 30~GHz channel, and much more sensitive than the WMAP K band.}
    \label{fig:noise-spectra}
\end{figure}

\begin{figure}
    \centering
    \includegraphics[width=0.49\textwidth]{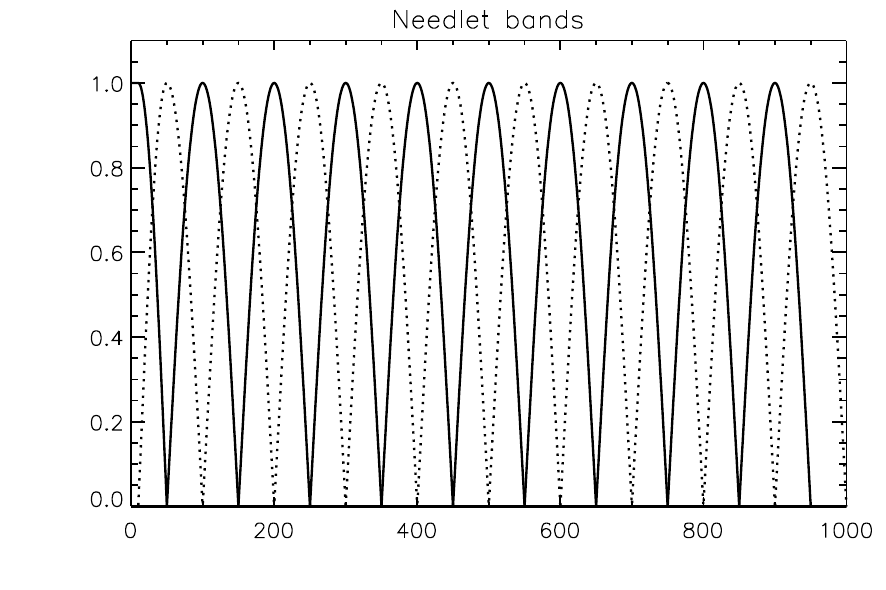}
    \includegraphics[width=0.49\textwidth]{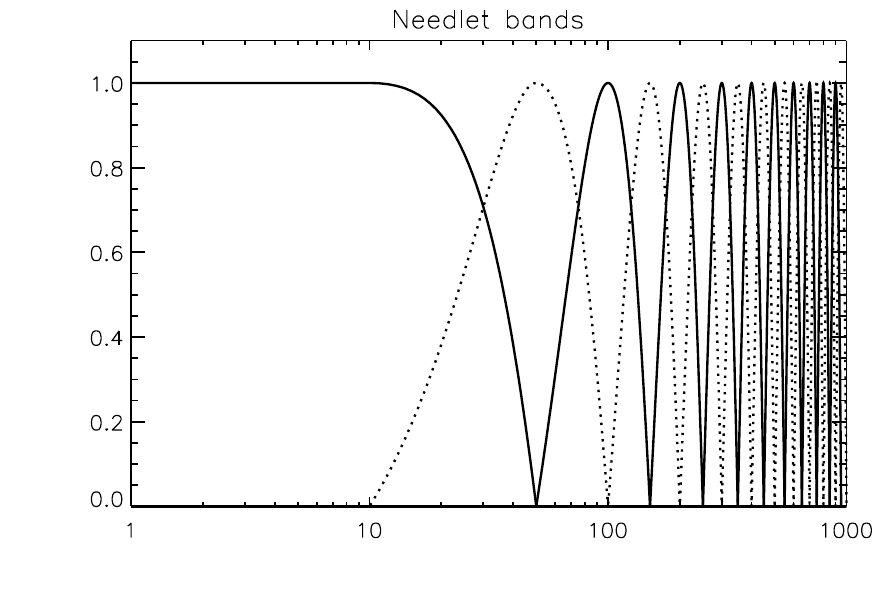}
    \caption{Cosine-shaped needlet bands $h_J(\ell)$ used in this work. As usual, those are normalised so that $\sum_J h_J(\ell)^2=1$ for a useful range of $\ell$. Here, this normalisation holds up to $\ell=950$. For higher $\ell$, the sum goes smoothly to zero at $\ell_{\rm max}=1000$ with a cosine-square half-arch.}
    \label{fig:needlet-bands}
\end{figure}

By reason of frequency-dependent angular resolution, it also depends on angular scale (Fig. \ref{fig:noise-spectra}). Finally, it depends on the Stokes parameter being considered, $Q$ or $U$. We hence propose to combine the observations using relative weights that depend on sky direction, harmonic mode, and polarisation orientation. This is done using a decomposition of the maps onto a needlet frame. The so-called ``needlets'' are a kind of wavelets defined on the sphere. For our purpose, needlet decomposition is performed by filtering the original maps in a set of harmonic windows $h_J(\ell)$, where $J$ indexes the scale and $\ell$ is the harmonic mode of spherical harmonics transforms. These needlet windows are normalised so that
\begin{equation}
    \sum_J h_J(\ell)^2 = 1
\end{equation}
over a useful range of $\ell$. Maps of needlet coefficients are obtained by an inverse Spherical Harmonic Transform (SHT) on the filtered harmonic coefficients. Map reconstruction from the needlets is done by filtering again each needlet coefficient map by the corresponding spectral window, co-addition in harmonic space, and an inverse SHT of the recombined $a_{\ell m}$ coefficients. 

Stokes parameters $Q$ and $U$ maps are defined with respect to a set of axes. They are not invariant with a transformation of those local reference axes. For this reason, they are not scalar quantities, but spin-2 fields. In CMB data analysis, it is customary to decompose polarisation maps into an alternate basis, the $E$ and $B$ polarisation modes, respectively scalar (even parity) and pseudo-scalar (odd parity). In previous work using needlets for polarisation, needlet windows have been used to filter $E$ and $B$ fields, which are then transformed into needlet coefficient maps for $E$ and $B$, processed, and then recombined. While this makes sense for the analysis of CMB maps, for which $E$ modes and $B$ modes are a natural decomposition, with $E$ modes outshining $B$ modes by orders of magnitude, it is not the case for maps of Galactic foreground components, for which $E$ and $B$ modes are of the same order of magnitude and have no special theoretical justification. On the contrary, the transformation of $Q$ and $U$ into $E$ and $B$ is non-local (while foreground emission is), and also sub-optimally smoothes the inhomogeneity of the noise level (the standard deviation is pixel-dependent). Moreover, the noise level is different in $Q$ and $U$ maps, as it depends on the orientation of the detectors as they scan through the sky pixels, a property that should be taken into account for the use of optimal weights for the production of low noise maps. 

In this work, noise-weighted combinations are performed pixel by pixel in needlet coefficient maps, but instead of working on needlet coefficients for $E$ and $B$ as done in previous work
\citep{2013MNRAS.435...18B,2023JCAP...06..034A,2022ApJ...940...68E}, we transform back $E$ and $B$ maps filtered by the spectral windows $h_J(\ell)$ into maps of $Q$ and $U$, make noise-weighted combinations of those ``special needlets'' in pixel space, transform them back into $E$ and $B$ $a_{\ell m}$s which are then filtered again by the spectral windows and co-added to produce $Q$ and $U$ synchrotron maps with near-optimal signal to noise ratio.

The specific data processing pipeline set up in this work is summarised as follows:
\begin{enumerate}
    \item Planck 30~GHz and 44~GHZ frequency maps, and WMAP K, Ka and Q maps are smoothed to 60$^\prime$ angular resolution in harmonic space;
\item CMB polarisation is subtracted from all frequency maps to minimise residual CMB in the maps; 
\item All maps are rescaled to equivalent synchrotron emission at the 30~GHz reference frequency, assuming a power law frequency scaling with spectral index $\beta_s = -3.1$;
\item We compute needlet coefficients by filtering $E$ and $B$ $a_{\ell m}$s using a set of 20 bands spanning the $\ell=2$--1000 range of harmonic modes, and producing corresponding maps of $Q$ and $U$ needlets at {\tt nside}=512;
\item We compute the noise level of each frequency channel in this needlet space, as a function of scale and sky pixel, and get inverse variance weighted averages for $Q$ and $U$ in each needlet band;
\item Those needlet coefficients are then used to reconstruct a single map of synchrotron $Q$ and $U$ polarisation at 60$^\prime$ angular resolution, which can be used as a template for polarised synchrotron emission at 30~GHz;
\item Two hundred maps of simulated noise for the end-product are obtained by co-adding WMAP and Planck noise simulations with the weights used in the production of the polarised synchrotron maps.
\end{enumerate}
Each of these steps is further detailed in the following sub-sections.

\subsection{Step 1: Map smoothing}
Original WMAP and Planck maps used in this work have angular resolutions ranging from about 27.9$^\prime$ (for the Planck LFI 44~GHz band, with the best resolution of the five) to about 52.8$^\prime$ (for the WMAP K band). As a first step, we transform original maps into spherical harmonics, and filter the map $a_{\ell m}$s by the ratio of the $B_\ell$ of a 60$^\prime$ beam and the $B_\ell$ of the original map. For WMAP we use tabulated beam $B_\ell$s provided with the WMAP DR5. For the Planck LFI, we assume the original beams are 33.1$^\prime$ and 27.9$^\prime$ for the 30~GHz and 44~GHz respectively. Small differences between those assumed values and the actual beams (which are known to be somewhat elliptical) have little impact on the final products, for which high-$\ell$ power is further suppressed.

The choice of 60$^\prime$ as the final resolution is motivated by an inspection of the power spectrum of the initial maps, which is noise dominated above values of $\ell$ ranging from $\ell=200$ at best (for near-full sky LFI 30~GHz maps) to $\ell=20$-30 at worst (for high Galactic latitude maps at 44~GHz and in the WMAP Q band. The whole pipeline has been run also with 50$^\prime$ and 40$^\prime$ target resolution, but the maps produced are noisier and contain little extra signal information in regions of faint emission.\footnote{Those maps can be made available upon request, but are not part of the default data products made available with this paper.}

\begin{figure}
    \centering
    \includegraphics[width=0.49\textwidth]{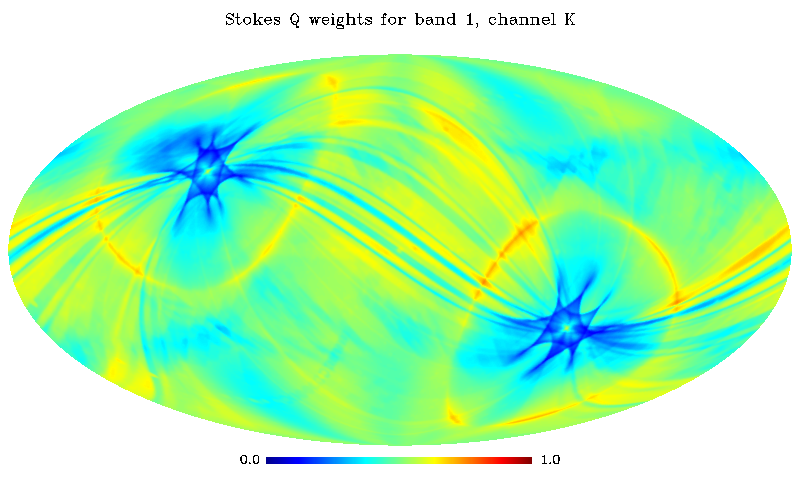}
    \includegraphics[width=0.49\textwidth]{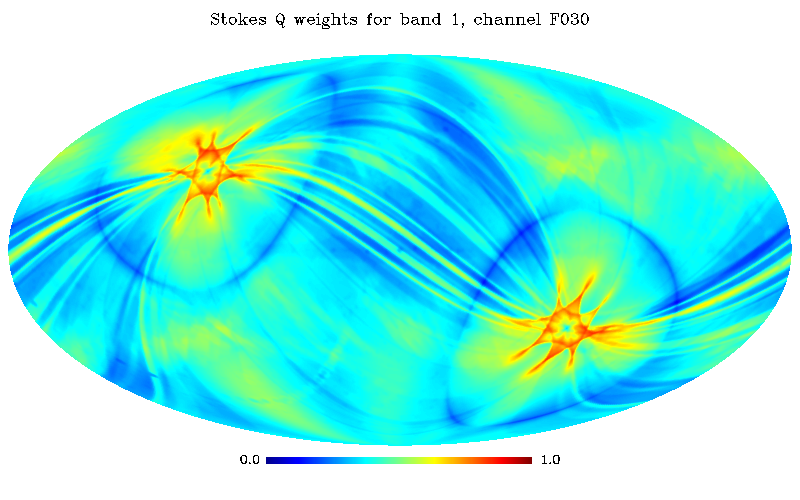}
    \includegraphics[width=0.49\textwidth]{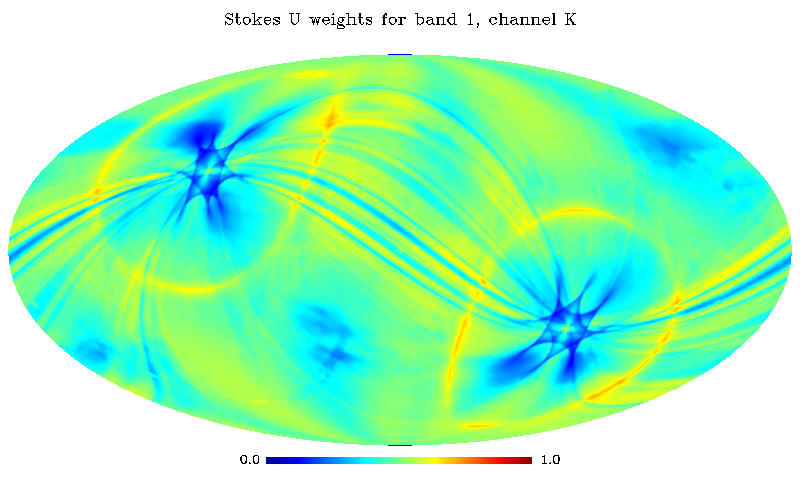}
    \includegraphics[width=0.49\textwidth]{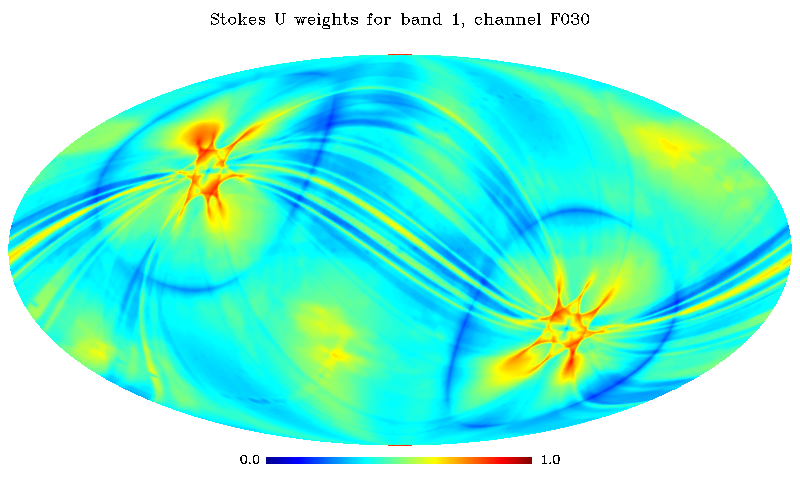}
    \caption{Weights used for the reconstruction of the synchrotron $Q$ (top two) and $U$ (bottom two) synchrotron polarisation maps in the first needlet band. Only the weights of the WMAP K and the Planck LFI 30~GHz maps are shown, as only those are significant for that scale (and add-up to almost 1). Note the differences between $Q$ and $U$ weights.}
    \label{fig:band-1-weights}
\end{figure}

\begin{figure}
    \centering
    \includegraphics[width=0.49\textwidth]{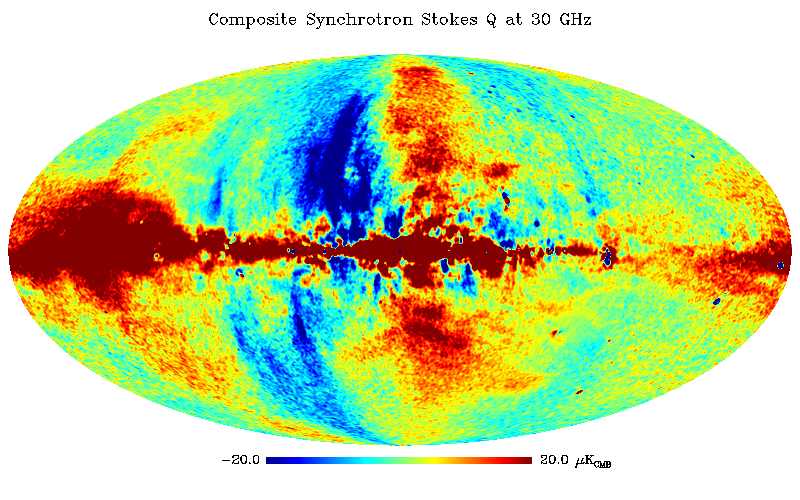}
    \includegraphics[width=0.49\textwidth]{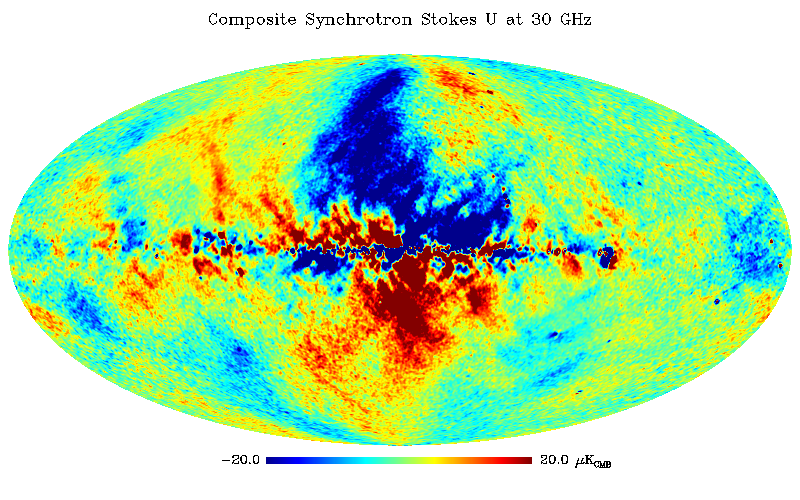}
    \caption{Stokes Q (top) and U (bottom) 30~GHz polarisation maps obtained with the pipeline described in this paper, at one degree angular resolution.}
    \label{fig:sync-QU-maps}
\end{figure}

\begin{figure}
    \centering
    \includegraphics[width=0.49\textwidth]{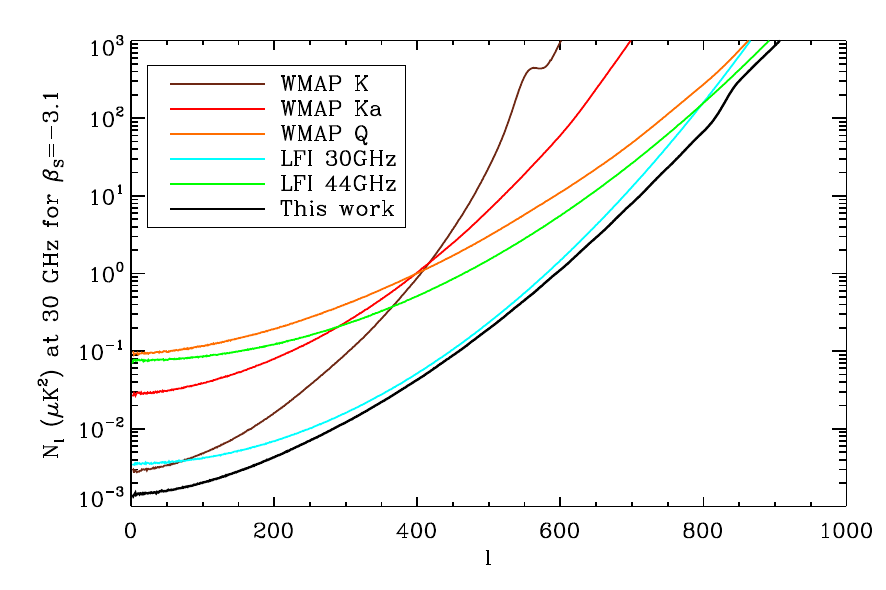}
    \includegraphics[width=0.49\textwidth]{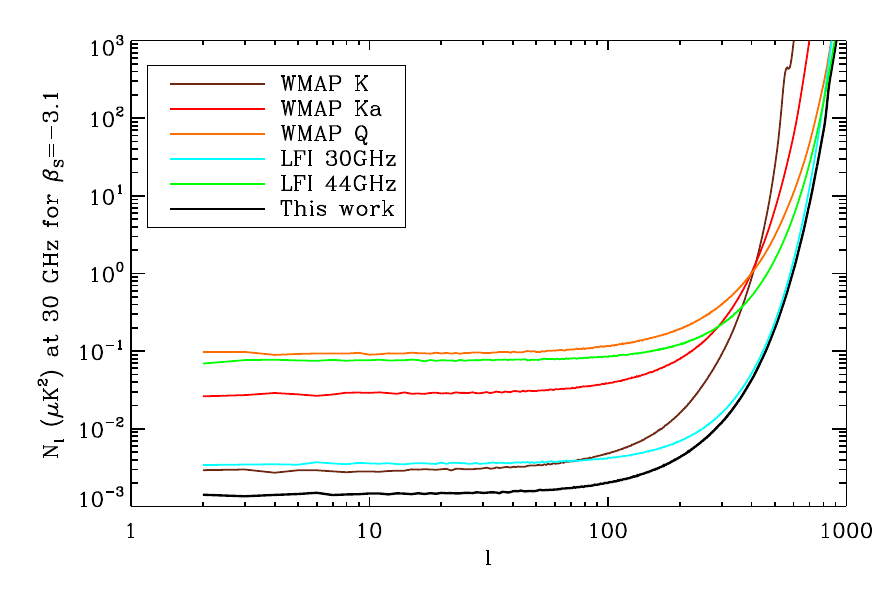}
    \caption{Beam-corrected noise spectra for the WMAP and Planck channels used in this work, re-scaled to equivalent 30~GHz noise assuming a synchrotron spectral index of $-3.1$, as in Fig.~\ref{fig:noise-spectra}. An extra black curve shows the noise spectrum of the final composite synchrotron map obtained in this work, which near-optimally combines all observations.}
    \label{fig:noise-spectra-final}
\end{figure}

\subsection{Step 2: CMB subtraction}

All of the maps used in this work contain some CMB polarisation. This CMB ``contamination'' is sub-dominant as compared to synchrotron emission and to noise contamination, but for the sake of optimality we subtract from each of the maps a Wiener-filtered version of the SMICA polarisation maps. Multivariate Wiener filtering is performed assuming theoretical CMB intensity and polarisation spectra corresponding to the best-fit Planck cosmological model. The multivariate spectrum of the noisy CMB maps is computed directly on the SMICA maps. Subtracting a Wiener-filtered CMB (rather than the SMICA CMB polarisation maps themselves) ensures that CMB subtraction does not add more noise than subtracts CMB. The residual noise in the maps is subdominant and, since it originates predominantly from HFI observations, is not significantly correlated with original noise in the WMAP and LFI channels. Its contribution, as well as that of the residual CMB, can safely be neglected in the computation of the weights to be used for the production of the synchrotron map.

We do not describe this step in more detail as it has almost no impact on the final products. Synchrotron $Q$ and $U$ maps have been produced with pipelines including or omitting CMB subtraction. Differences are imperceptible. Still, one should be aware that there is low-level residual CMB in the final products even for the pipeline that includes CMB subtraction (the difference between the Wiener filtered CMB and the original CMB is present in the maps). This residual can be characterised precisely if necessary for specific analyses. \footnote{Residuals are small, and for most uses can be safely neglected. In case of doubt contact the author. %The level of residuals depends on both the sky pixel and the angular scale, and there is no simple, single way to characterise those residuals that would be adapted a priori to any use of the data products.
}.

\subsection{Step 3: Rescaling to 30~GHz reference frequency}
After the two previous steps, we are left with five noisy synchrotron-dominated sky maps, all of which share the effective angular resolution of a 1-degree Gaussian beam over their common range of used $\ell$ values. For each of them, the useful $\ell$ range is restricted by the availability or by the accuracy of the instrumental effective beam window functions. We use $\ell_{\rm max}=[700,750,1000,850,1000]$ for WMAP K, Ka, Q, and LFI 30~GHz and 44~GHz respectively. 

We then use the WMAP and LFI bandpasses to integrate a single power law with spectral index of -3.1 (in K$_{\rm RJ}$ units) and use the corresponding synchrotron bandpass-integrated coefficients to rescale each map to equivalent synchrotron emission at 30~GHz (Table~\ref{tab:scalings}). This rescaling, by reason of variation of the synchrotron spectral index (or more generally, the synchrotron frequency scaling) across the sky, is slightly inaccurate. We evaluate the amplitude of the errors by computing the scaling coefficients for spectral indices ranging from -2.9 to -3.3. Differences of a few per cent at most across frequency channels guarantee that scaling errors are at most a few per cent of the total signal (and  in practice less, taking averaging between input maps on both sides of the reference frequency into account). These errors are below the noise over most of the sky, the only exception being regions of emission with signal to noise ratio of a few tens or more (in amplitude), for which there is no need to use the synchrotron maps produced in this paper rather than the original observations. We evaluate uncertainties more rigorously in Sec.~\ref{sec:results}.

\subsection{Step 4: Needlet coefficients}
CMB-corrected maps, smoothed to 60$^\prime$ resolution and scaled to 30~GHz synchrotron assuming a $\beta_s=-3.1$ spectral index are filtered in a set of cosine-shaped ``needlet bands'', shown in Fig.~\ref{fig:needlet-bands}. Those are chosen to be wide enough to allow for localisation on the scale of about a degree, allowing for adapting weights to map depth inhomogeneities on that scale. They are narrow enough for the signal to noise ratio of the various channels not to vary too much in each band. 

In practice, $Q$ and $U$ maps in each channel are transformed into $E$ and $B$ $a_{\ell m}$s, which are then multiplied by the windows $h_J(\ell)$, and transformed back using an inverse polarised SHT to produce a number $N_J$ of each of the $Q$ and $U$ needlet coefficient maps. 

\subsection{Step 5: Inverse noise variance weighting}
For each ``needlet'' map of $Q$ and $U$ Stokes parameters, we estimate the local noise variance integrated in the band. This is done by multiplying the original noise level in map space by a re-scaling factor that takes into account noise reduction in the band by reason of map smoothing (i.e. taking into account the ratio of the beam $B_\ell$s squared, integrated in the needlet band. We then co-add the needlet maps for the various channels with weights proportional to $1/\sigma^2$, normalised for unit response. This gives us 20 maps of needlet coefficients for each of $Q$ and $U$, which near-optimally combine the five original WMAP and LFI $Q$ and $U$ needlet coefficients for that scale into one single map of synchrotron emission needlet coefficients for each of $Q$ and $U$ at 30~GHz.

The relative weights for the various channels depend on the needlet scale (WMAP K and LFI 30~GHz dominate the weights on large scale, while other channels contribute at smaller scale), on the sky pixel, and on the Stokes parameter (weights are different for $Q$ and $U$, because the noise levels in each of those depend on the orientation of the scanning with respect to local polarisation reference axes.

For illustration purposes, weights for the WMAP K band and for the LFI 30~GHz channel, for the first needlet scale, are shown in Fig.~\ref{fig:band-1-weights}. In regions near the ecliptic poles, most of the weight is given to the LFI 30~GHz map, while over a large fraction of the sky, the WMAP K channel has a larger weight in the reconstruction. 

\subsection{Step 6: Coaddition}
Maps of needlet coefficients for $Q$ and $U$ are straightforwardly recombined into co-added maps as usual: polarised SHTs are performed to transform needlet $Q$ and $U$ maps into $E$ and $B$ $a_{\ell m}$s for each band. Those $a_{\ell m}$s are multiplied each by the appropriate needlet window $h_J(\ell)$. Those are then coadded before a final inverse SHT is performed to get final $Q$ and $U$ synchrotron maps with all scales co-added. 

\begin{figure}
    \centering
    \includegraphics[width=0.49\textwidth]{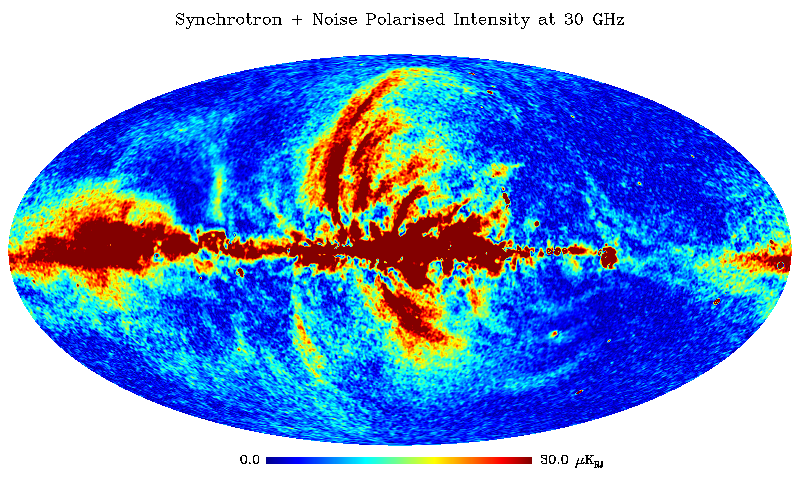}
    \includegraphics[width=0.49\textwidth]{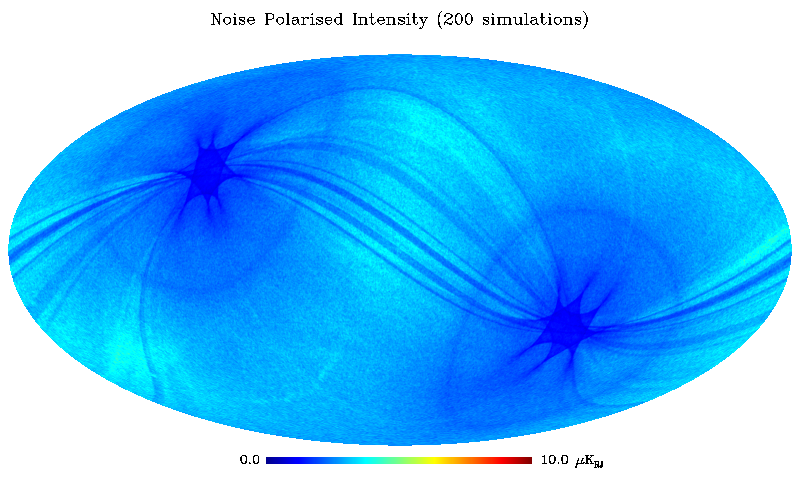}
    \includegraphics[width=0.49\textwidth]{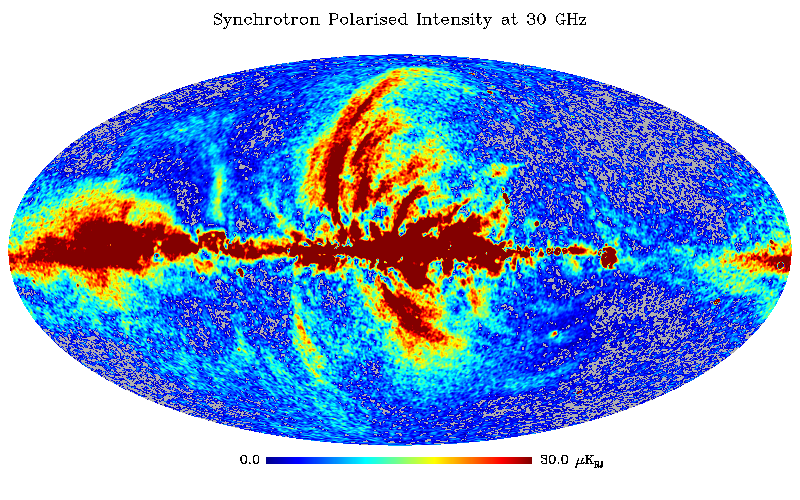}
    \caption{Synchrotron and noise polarised intensity as obtained in this work. Top: Total polarised intensity $P=\sqrt{Q^2+U^2}$ of the synchrotron map. Middle: Noise contribution, computed from 200 noise simulations. Increased map depth is visible both in regions of low Planck HFI noise near the ecliptic poles, and along the deeper integration rings of the WMAP scanning. Bottom: Synchrotron polarised intensity after de-biasing from the noise contribution. In the cleanest sky regions, pixels where $(Q^2+U^2) - (\sigma_Q^2+\sigma_U^2)$ is negative appear in grey.}
    \label{fig:final-noise-map}
\end{figure}

\begin{figure}
    \centering
    \includegraphics[width=0.49\textwidth]{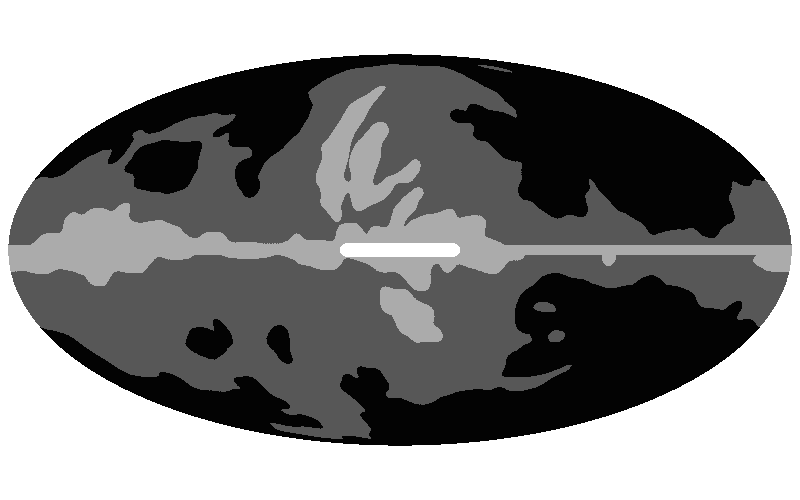}
    \includegraphics[width=0.49\textwidth]{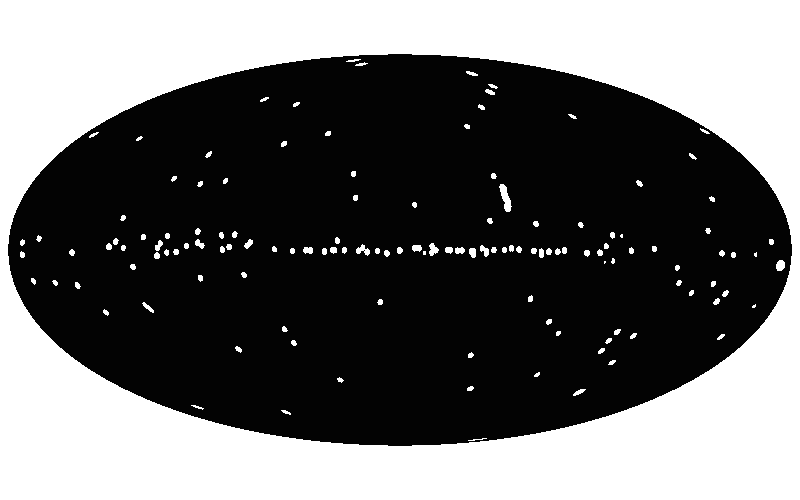}
    \caption{Masks used for computing power spectra shown in Fig.~\ref{fig:compare-cosmoglobe}. In the top panel, the low, medium, and high-amplitude regions of emission are shown in black, dark grey, and light grey respectively. The white region, located in the inner Galactic ridge, corresponds to the strongest region of polarised synchrotron emission, and is excluded for all power spectra computations. The bottom panel shows a mask for compact sources.}
    \label{fig:masks}
\end{figure}

\begin{figure}
    \centering
    \includegraphics[width=0.49\textwidth]{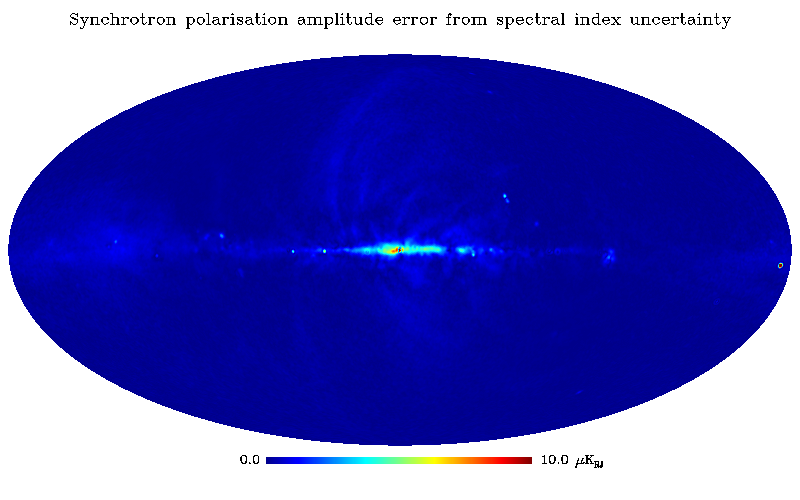}
    \includegraphics[width=0.49\textwidth]{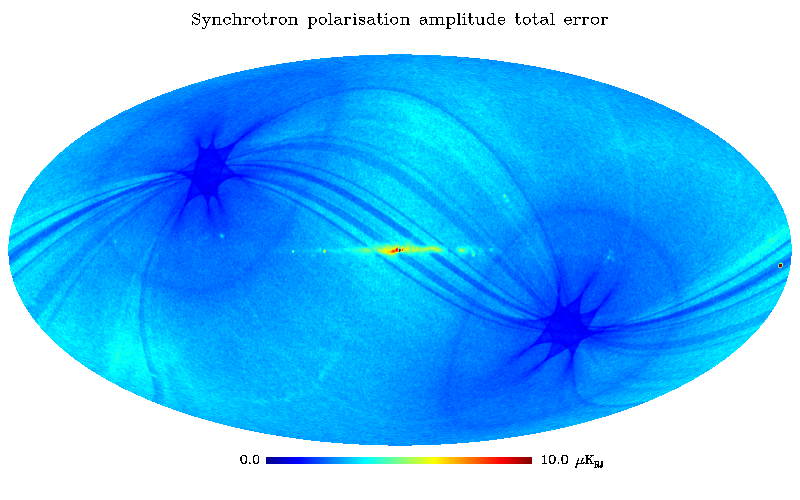}
    \caption{Impact of spectral index uncertainty on the polarised intensity maps. Top: Standard deviation of 50 maps obtained with $\beta_s$ drawn at random with a Gaussian distribution centered on $\beta_s=-3.1$ and with a standard deviation of 0.1. Bottom: Sum in quadrature of the above error and of the noise standard deviation.}
    \label{fig:specind-error}
\end{figure}

\begin{figure}
    \centering
\includegraphics[width=0.49\textwidth]{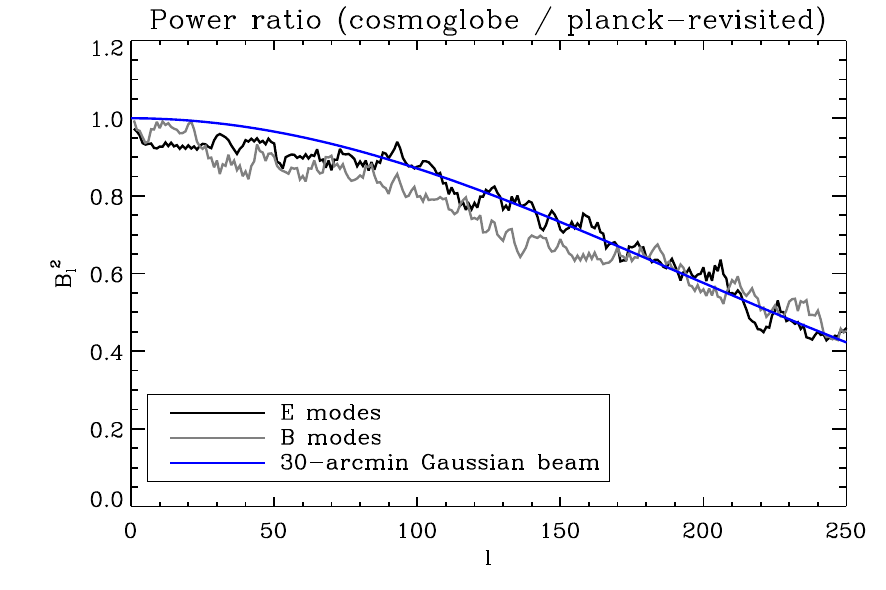}
    \caption{Ratio of cosmoglobe to Planck-Revisited $EE$ and $BB$ power spectra over the region of sky with highest emission.}
    \label{fig:CG_extra_smoothing}
\end{figure}

\begin{figure}[]
    \centering
    \includegraphics[width=0.49\textwidth]{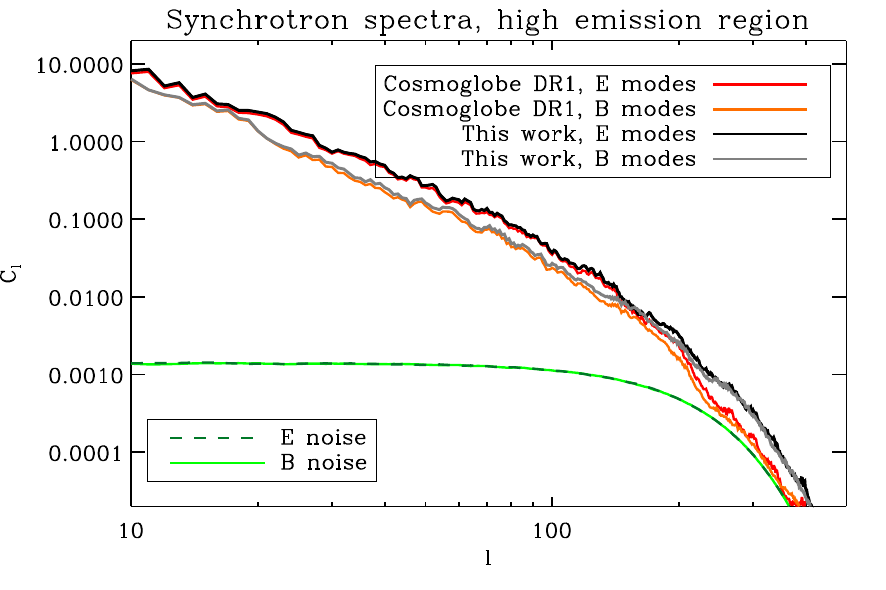}
    \includegraphics[width=0.49\textwidth]{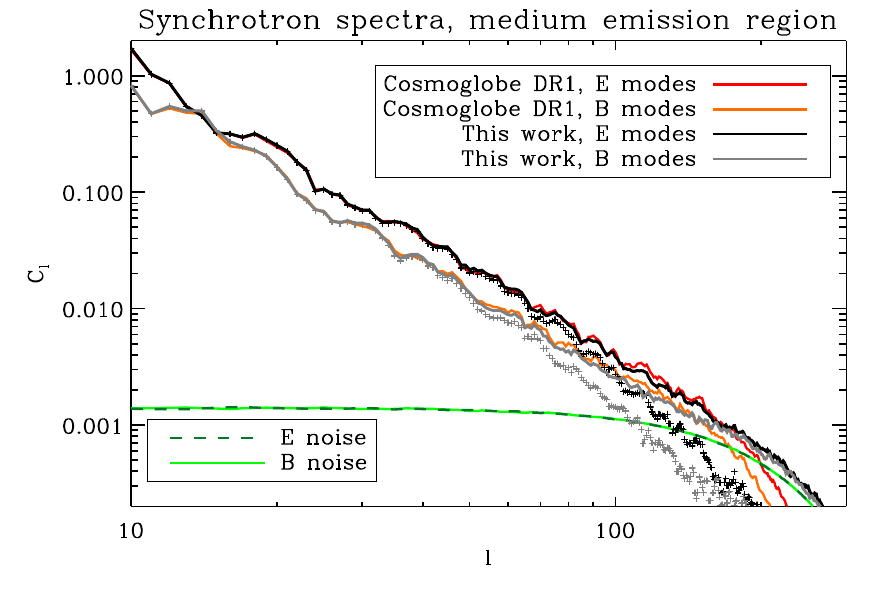}
    \includegraphics[width=0.49\textwidth]{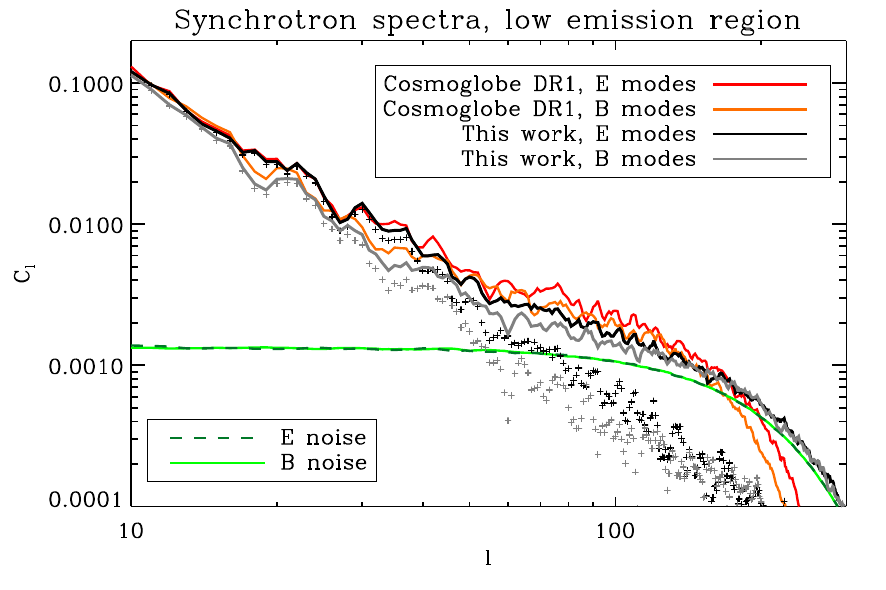}
    \caption{Power spectra of polarised synchrotron emission in three different sky regions, selected using the masks outlined in Fig.~\ref{fig:masks}. The data is plotted after smoothing the spectra with a $\Delta \ell=3$ window. In the high and medium emission regions (top and middle panels), spectra are strongly signal-dominated. There is excellent agreement between our maps and the cosmoglobe DR1 maps up to $\ell \simeq 30$, above which the cosmoglobe spectra are somewhat lower. At scales where signal dominates over noise, this discrepancy, increasing with increasing $\ell$, must be due to different effective beams. In the low emission region (bottom panel), of specific interest for future ground-based observations of CMB B-modes, the cosmoglobe polarisation maps have significantly higher power for $50 < \ell < 100$ than our maps. This is indicative of higher noise contamination, while the lower power of the cosmoglobe maps at higher $\ell$ likely indicates larger effective beam. For the medium and low contamination regions, we show the noise-debiased power spectra of our synchrotron polarisation maps with ``$+$''  data points. Noise spectra are computed from 200 simulations of WMAP and Planck LFI noise, propagated through the analysis pipeline.}
    \label{fig:compare-cosmoglobe}
\end{figure}

\subsection{Step 7: Noise simulations}
Synchrotron maps obtained following the previous steps comprise noise from WMAP and Planck observations, combined in a non-trivial way (non stationary, scale dependent, polarisation orientation dependent contribution from the various channels). This noise is not easily characterised analytically, as it is also correlated between pixels by reason of smoothing of the original observations. Two hundred simulated noise maps are obtained by projecting WMAP and Planck LFI simulated noise using the weights used for the various channels as a function of sky pixel and scale. For each simulation, random noise is generated using pixel-based covariance matrices for the $I$, $Q$ and $U$ noise provided by the WMAP and Planck collaborations, produced at the native {\tt nside} and angular resolution. These noise maps are then smoothed to 60$^\prime$ angular resolution, rescaled to 30~GHz using the synchrotron frequency scalings, decomposed into $Q$ and $U$ needlets, weighted and recombined into a single noise realisation. The statistics across the 200 realisations are sufficient to characterise noise contamination properties with a few per cent error, and global statistics (i.e. noise rms, spectrum) with sub percent accuracy.

\section{Results, discussion and data products}
\label{sec:results}

The main final data product is a pair of estimated $Q$ and $U$ synchrotron emission maps at 30~GHz, with a one degree Gaussian beam angular resolution. Maps of $Q$ and $U$ are shown in Fig.~\ref{fig:sync-QU-maps}. The co-addition of the original WMAP and Planck LFI observations using weights that depend on sky pixel and on angular scale for each of $Q$ and $U$ independently allows for significantly improved signal to noise ratio as compared to any of the original input maps. 

Figure~\ref{fig:noise-spectra-final} compares the $E$ and $B$ noise spectra of the final synchrotron polarisation obtained in this work to the original noise level of the input WMAP and Planck LFI channels (all re-scaled to 30~GHz assuming a synchrotron spectral index $\beta_s = -3.1$). As expected, the noise of our final synchrotron maps is lower than any of the noise spectra of the input maps used in this analysis.

Figure \ref{fig:final-noise-map} shows a map of synchrotron polarised amplitude, together with a map of noise standard deviation obtained from 200 simulated noise maps. Low-noise regions corresponding to both WMAP and Planck deeper integration zones are clearly visible in the standard deviation map, showing how the present method takes the best of the various observations used.  

No specific attempt has been made to subtract point source emission in the initial maps nor in the final polarisation maps. Well known bright radio sources can be identified in the final map. This is the case of supernova remnants in our own Galaxy such as Tau A and Cas A; of bright radio-galaxies such as Cen A, Fornax A and Virgo A; of bright distant quasars such as 3C273 and 3C279. A catalogue of polarised radio sources detected with Planck is available for identification and masking. Note that since the emissions from these sources do not necessarily scale with frequency with a power law with a -3.1 spectral index, their flux in the maps is not expected to correspond to their true 30~GHz emission. For convenience, a compact source mask (shown in Fig.~\ref{fig:masks}) is made available as part of the data products.

We anticipate that the present maps can be used as synchrotron  polarisation $Q$ and $U$ templates for foreground cleaning in upcoming CMB observations, in particular at few degree angular scales for detecting primordial B-modes of CMB polarisation. They will also be useful for making new models of Galactic foregrounds to be used in sky simulation tools such as the PSM~\citep{2013A&A...553A..96D} and the PySM~\citep{2017MNRAS.469.2821T,2021JOSS....6.3783Z}, to be used for the development of data analysis pipelines in preparation of future observations at millimeter to centimeter wavelengths. The good characterisation of the maps in terms of effective beam and the availability of 200 noise simulations will make it easy to combine the map with independent observations for yet-improved polarised synchrotron emission maps.

\subsection{Impact of the choice of a spectral index}
\label{sec:specind}

This work produces synchrotron polarisation $Q$ and $U$ maps assuming a fixed synchrotron spectral index $\beta_s=-3.1$ across the sky. We evaluate the impact of the variability of $\beta_s$, and of the uncertainty on its actual value, by repeating the analysis 50 times, with spectral indices drawn at random with a Gaussian distribution centered on $\beta_s=-3.1$ and with a standard deviation of 0.1. The standard deviation of the output polarised intensity map is shown in the top panel of Fig.~\ref{fig:specind-error}. As expected, this error is larger in the regions of higher total polarised emission, and is subdominant over most of the sky, except in the Galactic plane and towards bright compact sources.

\subsection{Comparison with other polarised synchrotron maps}
\label{sec:comparison}

Full-sky synchrotron polarization maps from Planck data alone have been obtained for the Planck 2018 data release using the SMICA \citep{2003MNRAS.346.1089D,2008ISTSP...2..735C} and the Commander \citep{2004ApJS..155..227E,2008ApJ...676...10E,2019A&A...627A..98S} component separation pipelines \citep{2020A&A...641A...4P}. In both cases however, only Planck data have been used, resulting in a noise penalty with respect to the composite polarised synchrotron polarisation maps presented here. Our ``Planck-revisited'' maps are substantially better in terms of noise contamination, and of characterisation of the effective angular resolution. As all data products are publicly available, we leave further comparison of our maps with those data products to the reader.

More interesting is the comparison with maps of synchrotron amplitude and spectral index obtained from a combined analysis of Planck and WMAP data recently been published by \cite{2023arXiv231013740W}, which are also available for download on a dedicated web site.\footnote{ cosmoglobe.uio.no/products/cosmoglobe-dr1.html  }
%\url{ cosmoglobe.uio.no/products/cosmoglobe-dr1.html } }

%\footnote{ \url{cosmoglobe.uio.no/products/cosmoglobe-dr1.html}} 
Those synchrotron polarisation $Q$ and $U$ maps are produced nominally at 60$^\prime$ angular resolution, but a smoothing prior of 30$^\prime$ applied during the fitting, and an additional one degree smoothing added when taking the mean over the Gibbs chain, are susceptible to further degrade the effective angular resolution of the cosmoglobe maps. 

We estimate the cosmoglobe map angular resolution by computing the ratio
\begin{equation}
    B_\ell^2 = \frac{C^{\rm CG}_{\ell,1 \times 2}}{C_\ell^{\rm PR} -N_\ell^{\rm PR}},
\end{equation}
where $C^{\rm CG}_{\ell,1 \times 2}$ is the cross-spectrum of the HM1 and HM2 cosmoglobe maps, and $C_\ell^{\rm PR} -N_\ell^{\rm PR}$ is the noise-debiased auto-spectrum of our Planck-Revisited map. The ratio, which is roughly compatible with extra $30^\prime$ smoothing in the cosmoglobe maps, is displayed in Fig.~\ref{fig:CG_extra_smoothing} for the most interesting $\ell$-range for the detection of primordial CMB B-modes.

The noise level in the total polarisation amplitude, 
\begin{equation}
    \sigma_P = \sqrt{\sigma_Q^2+\sigma_U^2},
\end{equation}
is 3.4~$\mu$K$_{\rm RJ}$ in the cosmoglobe DR1. 
For comparison, the noise standard deviation obtained with the present analysis is of about 2.78~$\mu$K$_{\rm RJ}$ at 60$^\prime$ resolution. 
When the map is further smoothed with a 30$^\prime$ beam to match the cosmoglobe map effective resolution, as detemined above, this noise standard deviation reduces to 2.32~$\mu$K$_{\rm RJ}$, 32\% lower than that of the cosmoglobe DR1 maps. The low noise contamination we achieve in the Planck-Revisited synchrotron map is mostly due to efficient co-addition of the input maps across scales, pixels and as a function of the Stokes parameter being considered, $Q$ or $U$.

Power spectra for three different sky regions, corresponding to low, medium, and high polarised synchrotron emission amplitude (using the masks displayed in Fig.~\ref{fig:masks}), are shown in  Fig.~\ref{fig:compare-cosmoglobe}, together with power spectra computed from the cosmoglobe DR1 data products. For each region, the mask selecting the level of contamination is apodised with a 2-degree cosine-squared transition, and point sources are masked using the mask shown in the bottom panel of Fig.~\ref{fig:masks}, apodised with a 1-degree cosine-squared transition. 

Overall, there is good agreement between the two data products at low $\ell$. For intermediate $\ell$, the cosmoglobe synchrotron maps are noisier. At high $\ell$, the overall power in the cosmoglobe maps is lower, indicative of extra smoothing, as already determined above.

\subsection{Data products}

``Planck Revisited'' data products are made available on a dedicated website.\footnote{\url{https://portal.nersc.gov/project/cmb/Planck_Revisited}} Data products specific to this work comprise:
\begin{enumerate}
    \item Maps of synchrotron $Q$ and $U$ Stokes parameters as obtained with the method described in Sec.~\ref{sec:method}, at 60$^\prime$ resolution (maps at higher resolution can be made available by the author upon request);
    \item Masks dividing the sky in three regions of different synchrotron polarised amplitude (high, medium, and low polarised synchrotron emission);
    \item A mask for strong polarised compact sources, which can be completed if necessary using the Planck catalogue of  non-thermal source catalogue \citep{2018A&A...619A..94P};
    \item Maps of simulated noise for the 60$^\prime$ resolution synchrotron $Q$ and $U$ polarisation data products (additional simulations can be made available if necessary).
\end{enumerate}

No attempt has been made to correct for residual systematics in the maps, nor fit for the scaling of synchrotron emission as a function of sky direction. 
For these reasons, caution must be used when exploiting our maps in regions of strong emission, where the calibration uncertainty due to scaling with frequency, as well as potential intensity to polarisation leakage, can generate errors larger than the map noise. At very low $\ell$ (e.g. below $\ell=20$), systematics in the original data products are expected to project onto our synchrotron maps. However, in regions where the polarised synchrotron emission is low and for $\ell$-ranging from 20 to 100, our maps are the most accurate full-sky tracers of polarised synchrotron emission to date. In particular, they can be used as polarised synchrotron templates for upcoming searches, in low-foreground sky regions, of a recombination peak (around $\ell=80$) in CMB primordial B-modes. 

\section{Conclusion}
\label{sec:conclusion}

In this paper, we have applied a simple, yet effective pipeline to combine WMAP K, Ka and Q maps and Planck LFI 30~GHz and 44~GHz maps into a single set of $Q$ and $U$ synchrotron polarisation maps at 30~GHz. The effective angular resolution of the maps is that of a 60$^\prime$ Gaussian beam, with a smooth cosine-square-shaped apodisation from $\ell=950$ to $\ell=1000$. Additive errors from instrumental noise are evaluated with two hundred noise simulations propagated through the analysis pipeline. Multiplicative errors from the assumption of a uniform synchrotron emission law with $\beta_s=-3.1$, evaluated with a Monte-Carlo pipeline over a distribution of spectral indices, are evaluated and shown to be sub-dominant over most of the sky. 

The maps and noise simulations are made available to the scientific community on a dedicated data server, accessible from the web. 

\begin{acknowledgements} 
Some of the results in this paper have been derived using the HEALPix pixellisation scheme \citep{2005ApJ...622..759G}. The author thanks Duncan Watts for useful discussions regarding comparison with the cosmoglobe analysis and DR1 data products.
\end{acknowledgements}

\bibliographystyle{aa} 
\bibliography{biblio} 
%\bibliography{biblio,Planck_bib} 

\end{document}